\documentclass[conference, 11pt]{IEEEtran}

\IEEEoverridecommandlockouts   

\usepackage{dsfont}
\usepackage[T1]{fontenc}
\usepackage{graphicx, stfloats} 
\usepackage{svg}
\usepackage{geometry}
\usepackage{standalone}
\usepackage{hyperref}
\usepackage{pgfplots}
\usepgfplotslibrary{groupplots}
\usepackage{booktabs}
\usepackage{subcaption}
\usepackage{amssymb, amsmath}
\usepackage{algorithm}
\usepackage{algpseudocode}
\usepackage{makecell}

\usepackage{diagbox, multirow} 
\usepackage{color,soul} 
\usepackage{siunitx}
\usepackage{xurl}

\newcommand{\rev}[1]{{\color{black}#1}}

\definecolor{vandeusen}{RGB}{73,92,111}
\definecolor{cordovan}{RGB}{152,68,71}
\definecolor{alizarin}{rgb}{0.82, 0.1, 0.26}
\definecolor{azure}{rgb}{0.0, 0.5, 1.0}

\pgfplotsset{compat=1.16}
\usepgfplotslibrary{statistics}
\usetikzlibrary{patterns}

\definecolor{barVI}{RGB}{31,119,180}
\definecolor{barEqMCTS}{RGB}{44,160,44}
\definecolor{barMCTS}{RGB}{23,190,207}
\definecolor{barEqGreedy}{RGB}{148,103,189}
\definecolor{barMkGreedy}{RGB}{255,127,14}
\definecolor{barRandom}{RGB}{214,39,40}
\definecolor{barUS}{RGB}{31,119,180}
\definecolor{barID}{RGB}{255,127,14}

\pgfplotsset{
  resultbar/.style={
    ybar,
    width=\linewidth,
    height=5.2cm,
    ymajorgrids=true,
    grid style={gray!25},
    tick align=outside,
    tick pos=left,
    axis line style={gray!60},
    xtick=data,
    x tick label style={font=\footnotesize, rotate=30, anchor=east},
    y tick label style={font=\footnotesize},
    label style={font=\small},
    title style={font=\small\bfseries, yshift=-2pt},
    error bars/y dir=both,
    error bars/y explicit,
    error bars/error bar style={gray!55, line width=0.6pt},
    enlarge x limits=0.12,
    every axis plot/.append style={draw=black!55, line width=0.4pt},
  },
  groupedbar/.style={
    ybar,
    width=\linewidth,
    height=5.4cm,
    ymajorgrids=true,
    grid style={gray!25},
    tick align=outside,
    tick pos=left,
    axis line style={gray!60},
    xtick=data,
    x tick label style={font=\footnotesize},
    y tick label style={font=\footnotesize},
    label style={font=\small},
    title style={font=\small\bfseries, yshift=-2pt},
    legend style={font=\footnotesize, draw=gray!50, at={(0.5,-0.28)}, anchor=north, legend columns=2},
    enlarge x limits=0.35,
    every axis plot/.append style={draw=black!55, line width=0.4pt},
  },
}

\title{Comparing Socially-Equitable Renewable Energy Budget Allocation MDP Policies \\in Mature and Emerging Economies}

\begin{document}

\author{
\IEEEauthorblockN{Riya Kinnarkar}
\IEEEauthorblockA{
\textit{Management \& Technology} \\
\textit{University of Pennsylvania} \\ \text{Philadelphia, PA, USA}}
\\
\IEEEauthorblockN{Yan Pratama Akhra}
\IEEEauthorblockA{\textit{Industrial and Systems Engineering} \\
\textit{Institute Technology of Sepuluh Nopember (ITS)} \\
\text{Surabaya, Indonesia}}
\and
\IEEEauthorblockN{Mansur M. Arief}
\IEEEauthorblockA{\textit{Industrial and Systems Engineering} \\
\textit{King Fahd University of Petroleum} \\ \textit{and Minerals (KFUPM)}, Saudi Arabia}
\\
\IEEEauthorblockN{Dino Arla}
\IEEEauthorblockA{\textit{PT PLN Persero}\\
\text{Indramayu, Indonesia}
}
}

\maketitle

\begin{abstract}
Equitable renewable-energy planning is a sequential decision problem, but the decision variables available to a public planner differ sharply between mature and emerging economies. In the former the government largely builds generation, while in the latter it steers private investment through incentives and quotas. We formulate socially-equitable renewable-energy budget allocation as a Markov Decision Process (MDP) and, using a single problem-agnostic solver interface, compare the same policies across the two settings: eight U.S. cities (a mature economy) and West Java, Indonesia (an emerging economy). The results show that across both settings, a receding-horizon value-iteration policy dominates. In the U.S., it reaches 66\% renewable penetration while cutting the underserved low-income population by 96\% versus a random baseline. In West Java it closes the low-access gap while crowding in \rev{about seven times the private capital of a naive market-chasing heuristic}. More interestingly, \rev{this heuristic}, which is mildly sub-optimal in the U.S., could yield catastrophic outcomes in Indonesia, by underserving every low-access region, because chasing attractive markets and serving the underserved \rev{diverge} once the planner acts through private developers.
\end{abstract}

\begin{IEEEkeywords}
Renewable energy, social equity, Markov Decision Process, grid optimization, energy justice
\end{IEEEkeywords}

\section{Introduction}

The transition to renewable energy spans both technical and societal considerations. Grids designed for dispatchable fossil generation struggle to absorb variable wind and solar. While renewable investment has surged, grid infrastructure funding has lagged, leaving thousands of gigawatts of projects awaiting connection~\cite{iea_2023_grids}. At the same time, the burdens of an unreliable grid fall disproportionately on low-income and socially vulnerable communities: U.S. households in poverty spend more than twice the average share of income on energy~\cite{pearl_elevate_2024_energy_gap}. After major storms, for example, communities lower on the CDC social vulnerability index wait significantly longer for power to return~\cite{ji_ganz_2024_power_outages}. Yet, most renewable-energy planning tools optimize cost and reliability while treating equity as an add-on. Embedding social equity directly into the planner's objective is therefore both a modeling and a policy need.

The crux of our study is that the public levers for driving an equitable renewable transition differ fundamentally between \emph{mature} and \emph{emerging} economies. In a mature economy such as the U.S., a public planner can directly finance and build generation under a capital budget. In an emerging economy such as Indonesia, most renewable capacity is built by private companies and independent power producers selling to the state utility. To that end,  the government's lever is not construction but \emph{incentives and quotas} that steer where private capital flows. A planning model\rev{, and the policy insights drawn from it,} may not survive this shift. We therefore formulate socially-equitable budget allocation as a Markov Decision Process (MDP) and evaluate the \emph{same} set of policies on two settings: eight U.S. cities (the planner builds) and West Java, Indonesia (the planner incentivizes). We compare \rev{receding-horizon} value iteration~\cite{bellman1958dynamic}, Monte Carlo Tree Search (MCTS)~\cite{browne2012survey}, and equity- and market-oriented heuristics, and cross-deploy each policy's inductive bias from one setting to the other. 

Our contribution is as follows. First, we build \rev{an} MDP model of socially-equitable renewable-energy budget allocation that \rev{extends our earlier U.S. model~\cite{kinnarkar2025optimized} and} is applicable to both mature and emerging economies. Second, we compare \rev{MDP-based} policies across the two settings that differ in the planner's lever\rev{,} eight U.S. cities (a mature economy) and West Java, Indonesia (an emerging economy)\rev{,} to study the robustness of the outcomes and identify modes of failure \rev{from both} technical and societal perspectives. The comparison reveals that while a value-iteration policy and an equity-targeting objective are robust across both contexts, a market-chasing heuristic that is merely \rev{sub-optimal} when the planner builds becomes actively harmful when the planner incentivizes\rev{, which is a risk that neither setting reveals on its own}.

\section{Related Work}\label{sec:related_work}

Renewable-energy planning has traditionally been framed as a cost-and-reliability optimization. Long-horizon capacity-expansion tools such as ReEDS~\cite{nrel_reeds} simulate grid growth to guide investment. A growing body of evidence shows that higher renewable penetration correlates with fewer and shorter outages \rev{and not with} a more fragile grid~\cite{nieman_2025_renewables_grid}. Such models excel at aggregate trends but assume fixed conditions and rarely capture the distributional outcomes that matter for equity. In our view, this omission is consequential, as outages fall hardest on vulnerable communities\rev{, as} socially vulnerable neighborhoods and lower-income census tracts endure markedly longer losses after extreme weather~\cite{flores_casey_2024_power_outages, coleman2023energy}\rev{. Reviews} of data-driven grid resilience \rev{nevertheless} find that most methods still omit an explicit equity objective~\cite{zhao2024optimizing}, motivating frameworks that encode disparity in the decision process itself.

The MDP is a natural framework for handling stochastic, sequential decisions and \rev{admits} both exact~\cite{bellman1958dynamic} and sampling-based~\cite{browne2012survey} solvers. MDPs and related sequential methods have been applied to photovoltaic voltage regulation~\cite{el2021fully}, community storage for social welfare~\cite{deng2020community}, uncertainty-aware microgrid dispatch~\cite{wang2025energy}, household technology adoption~\cite{krisnawati2025qlearning}, and machine-learning-augmented system design~\cite{sanjuan2024ml}. These earlier studies mostly optimized technical objectives. In this study, we instead add another layer of social equity directly in the reward.

The social equity aspect and the corresponding policy levers in renewable energy planning and decision making \rev{are} an active area of research. \rev{Energy-justice research asks how the benefits and costs of energy services are distributed and who takes part in the decisions~\cite{sovacool2015energy}.} Earlier studies have \rev{also examined} the policy levers that emerging economies use to drive clean-energy investment\rev{, including} green-finance instruments that improve energy accessibility~\cite{gao2025green}, government guidance funds that raise renewable-firm productivity~\cite{xie2026government}, and quota mechanisms that shape clean-energy investment decisions~\cite{chen2026carbonquota}. \rev{Empirical work on Indonesia links slow private investment in renewables to regulatory barriers, a monopolized power market and changing pricing rules~\cite{burke2019overcoming, maulidia2019rethinking}, and finds that private financing of rural electrification favors large on-grid projects over small distributed ones~\cite{setyowati2021mitigating}. A core finding of this literature is} that, where the state regulates \rev{and does not build}, the design of incentives and quotas\rev{, and not direct construction,} determines outcomes. \rev{These two research directions, however, rarely converge}: sequential-decision models are typically posed for a single build-centric planner, while incentive-design studies rarely cast the problem as a sequential policy over heterogeneous regions. Our contribution bridges them by evaluating the \emph{same} equity-aware MDP policies across both institutional regimes and, crucially, by \emph{cross-deploying} each policy from one regime to the other to expose which insights transfer and what unintended failures and risks could emerge.

\section{MDP Formulation}\label{sec:framework}

We model equitable energy allocation as an MDP $(\mathcal{S}, \mathcal{A}, T, R, \gamma)$ over a set of $n$ regions $\mathcal{I} = \{1,\dots,n\}$.

\textbf{State.} The state at time $t$ is given by $s_t = (b, [d_i, r_i, n_i, p_i, I_i]_{i \in \mathcal{I}}),$ where $b$ is the remaining budget and, for region $i$, $d_i$ is energy demand, $r_i$ and $n_i$ are renewable (RE) and non-renewable (NRE) supply, $p_i$ is population, and $I_i \in \{0,1\}$ is an income indicator ($0$ = low income). Demand is stochastic, $d_i \sim \mathcal{N}(\mu_i, \sigma_i^2)$.

\textbf{Action.} At each step the planner adds or removes an RE or NRE facility in a chosen region $i$ (a capacity increment $\Delta s_r$ or $\Delta s_n$), or does nothing ($a_0$). Each build/retire action debits the budget by the corresponding capital cost ($c_{a,r}, c_{a,n}, c_{r,r}, c_{r,n}$) plus per-unit operating cost $c_{o,\cdot,i}$.

\textbf{Transition.} The dynamics factor into a \emph{deterministic} control component and a \emph{stochastic} exogenous component. Given action $a_t$ acting on region $i$, the capacity and budget update deterministically:
\begin{align}
r_i' &= r_i \pm \Delta s_r \ (\text{RE})\\
n_i' &= n_i \pm \Delta s_n \ (\text{NRE}), \\
b' &= b - c(a_t) - \textstyle\sum_{j} c_{o,\cdot,j}\, x_j',
\end{align}
where $c(a_t)$ is the capital cost of $a_t$ and the last term is total operating cost over installed capacity $x_j'$; all other regions are unchanged. The stochasticity enters through demand: at every step each region's demand is resampled, $d_i' \sim \mathcal{N}(\mu_i, \sigma_i^2)$, so the planner commits capacity before demand is realized and must be robust to demand fluctuation \rev{instead of} optimizing against a fixed forecast. The transition kernel is thus 
\begin{equation}
    T(s' \mid s, a) = \rev{\mathds{1}}(r',n',b'),\prod_{i}\mathcal{N}(d_i';\mu_i,\sigma_i^2),
\end{equation} 
i.e., a Dirac mass on the controllable variables times an independent Gaussian on demand. An episode terminates after a fixed horizon of $H$ decisions or when the budget is exhausted. The discount $\gamma$ then trades off near- versus long-term coverage. This structure keeps the model tractable for exact solvers on a discretized budget grid while preserving the demand uncertainty that makes purely myopic ``fill-the-largest-deficit'' rules fragile.

\textbf{Reward.} The reward balances three competing objectives, evaluated on the resulting state $s_{t+1}$ (the standard $R(s,a,s')$ convention, so an investment is credited in the step it is made). Let $\kappa(a_t)$ be the expenditure of $a_t$, $P_0 = \sum_{i: I_i=0} p_i$, and $P = \sum_{i} p_i$:
\begin{align}
R(s_t, a_t) &= w_1 \kappa(a_t) + \frac{w_2}{P_0}\sum_{i: I_i = 0} p_i \min\!\Big(1, \tfrac{r_i' + n_i'}{d_i'}\Big) \nonumber \\
&\quad + \frac{w_3}{P}\sum_{i \in \mathcal{I}} p_i \min\!\Big(1, \tfrac{r_i'}{d_i'}\Big),
\end{align}
with $w_1 = -0.02$, $w_2 = 200$, $w_3 = 80$ (Table~\ref{tab:experiment_params}) and $\gamma = 0.95$. The three terms are cost, the population-weighted \emph{coverage served} among low-income regions (equity), and the population-weighted renewable coverage (decarbonization). Rewarding coverage \emph{served} \rev{instead of} penalizing shortfall keeps the objective positive without changing the policy ranking, and charging the budget through $\kappa(a_t)$ \rev{instead of} crediting the remaining budget each step prevents the degenerate ``do nothing to preserve cash'' optimum.

\section{Numerical Experiments}\label{sec:experiment}

We study two cases that share the MDP structure of Section~\ref{sec:framework} and the \emph{same} solver code, but differ in the planner's lever: a mature economy where the planner builds generation (Case~1) and an emerging economy where the planner steers private capital through incentives and quotas (Case~2). Results for both are reported together in Section~\ref{sec:discussion}.

\textbf{Case 1: U.S. Model.} We instantiate the MDP on eight major U.S. cities (Atlanta, New York, Houston, Phoenix, Denver, Memphis, Seattle, San Antonio), chosen for diverse demographics, incomes, and energy mixes, using synthetic-but-plausible data (a city is ``low-income'' if $\geq 25\%$ of its population is low-income). The planner starts with a \$3{,}000M budget and makes 20 sequential decisions; costs, capacity increments, and reward weights are in Table~\ref{tab:experiment_params}.

\begin{table}[!htbp]
\centering
\caption{Case 1 Experiment Parameters (U.S. Model)}
\label{tab:experiment_params}
{\footnotesize
\begin{tabular*}{\linewidth}{@{\extracolsep{\fill}}ll@{}}
\toprule
\textbf{Parameter} & \textbf{Value} \\
\midrule
Initial budget & \$3{,}000M \\
Add RE / NRE cost ($c_{a,r}, c_{a,n}$) & \$180M / \$120M \\
Remove RE / NRE cost ($c_{r,r}, c_{r,n}$) & \$120M / \$180M \\
RE / NRE operating cost ($c_{o,\cdot,i}$) & \$8--14 / \$35--48 per MWh \\
Supply increment ($\Delta s_r, \Delta s_n$) & 10 GW \\
Horizon / discount ($\gamma$) & 20 decisions / 0.95 \\
Weights ($w_1, w_2, w_3$) & $-0.02$, $200$, $80$ \\
\bottomrule
\end{tabular*}
}
\end{table}

\begin{table}[!htbp]
    \centering
    \caption{Case 2 Parameters (West Java Model)}
    \label{tab:indonesia_params}
    {\footnotesize
    \begin{tabular*}{\linewidth}{@{\extracolsep{\fill}}ll@{}}
    \toprule
    \textbf{Parameter} & \textbf{Value} \\
    \midrule
    Initial incentive budget & IDR 1{,}000 bn \\
    Incentive / off-grid / quota cost & 40 / 90 / 5 (IDR bn) \\
    Incentive / off-grid capacity & 20 / 18 GW \\
    Quota release per action & 20 GW \\
    Max private leverage ($\times$ incentive) & 3.0 \\
    Cost weight ($w_1$) & $-0.02$ \\
    Access-coverage weight ($w_2$) & $180$ \\
    RE-coverage weight ($w_3$) & $70$ \\
    Leverage / deadweight weight & $+25$ / $-20$ \\
    Horizon / discount ($\gamma$) & 20 decisions / 0.95 \\
    \bottomrule
    \end{tabular*}
    }
    \end{table}

\textbf{Case 2: West Java Model.} In many emerging economies the government does not build but \emph{regulates and finances}: in West Java (Jawa Barat), most renewable capacity is built by private developers and IPPs selling to the state utility PLN. We recast the MDP accordingly. The \textbf{action space} shifts from ``what to build'' to ``where to direct an incentive or quota''\rev{: the planner can} issue an on-grid incentive (realized by private capital), fund an off-grid project directly, or release on-grid quota. On-grid deployment is capped by a per-region PLN \emph{quota} (post net-metering, based on local regulations), while off-grid projects serve remote regencies (Sukabumi, Garut, Tasikmalaya) where grid extension is costly. A \textbf{leverage} term credits \emph{additional} private capital drawn into regions the market ignores and penalizes \emph{deadweight} subsidies to already-attractive clusters~\cite{gao2025green,xie2026government}, and a per-cluster \textbf{quota constraint} follows the multi-year cycle~\cite{chen2026carbonquota}. The reward keeps the same coverage-served structure, now over low-\emph{access} regions, plus the leverage term. The experiment parameters are summarized in Table~\ref{tab:indonesia_params}.

\begin{table*}[!htbp]
    \centering
    \caption{Case 1 Policy Comparison\rev{:} U.S. Model}
    \label{tab:performance}
    {\footnotesize
    \begin{tabular*}{\linewidth}{@{\extracolsep{\fill}}lccccc@{}}
    \toprule
    \textbf{Policy} & \textbf{Reward} & \textbf{RE \%} & \textbf{Budget} & \textbf{Low-Inc} & \textbf{Low-Inc} \\
    & \textbf{$\uparrow$} & \textbf{$\uparrow$} & \textbf{(\$M)} & \textbf{Cities $\downarrow$} & \textbf{Pop (M) $\downarrow$} \\
    \midrule
    Value Iteration    & \textbf{4319 $\pm$ 64}  & 66.3 $\pm$ 1.6   & 2962 $\pm$ 25  & 1.3 $\pm$ 0.5    & \textbf{0.22 $\pm$ 0.08} \\
    Equity-MCTS        & 3960 $\pm$ 89           & 50.5 $\pm$ 4.1    & 1584 $\pm$ 50  & 3.5 $\pm$ 0.6    & 1.62 $\pm$ 0.32 \\
    MCTS               & 3933 $\pm$ 86           & 43.2 $\pm$ 4.2    & 2050 $\pm$ 50  & 3.0 $\pm$ 0.7    & 1.29 $\pm$ 0.31 \\
    Equity-Greedy      & 3921 $\pm$ 68           & 67.5 $\pm$ 0.6    & 2916 $\pm$ 59  & 2.4 $\pm$ 1.0    & 0.53 $\pm$ 0.45 \\
    Market-Greedy      & 3344 $\pm$ 59           & 24.0 $\pm$ 0.0    & 2430 $\pm$ 95  & 2.3 $\pm$ 1.1    & 0.47 $\pm$ 0.42 \\
    Random             & 2439 $\pm$ 353          & 45.9 $\pm$ 6.9    & 2873 $\pm$ 206 & 4.4 $\pm$ 0.8    & 5.61 $\pm$ 1.53 \\
    \bottomrule
    \end{tabular*}
    }
    \end{table*}
    
    \begin{table*}[!htbp]
    \centering
    \caption{Case 2 Policy Comparison\rev{:} West Java Model}
    \label{tab:indonesia}
    {\footnotesize
    \begin{tabular*}{\linewidth}{@{\extracolsep{\fill}}lccccc@{}}
    \toprule
    \textbf{Policy} & \textbf{Reward} & \textbf{RE \%} & \textbf{Budget} & \textbf{Low-Acc} & \textbf{Leverage} \\
    & \textbf{$\uparrow$} & \textbf{$\uparrow$} & \textbf{(IDR, billions)} & \textbf{Pop (M) $\downarrow$} & \textbf{$\uparrow$} \\
    \midrule
    Value Iteration & \textbf{3549 $\pm$ 40} & 52.4 $\pm$ 0.6 & 909 $\pm$ 43 & \textbf{0.00 $\pm$ 0.00} & 479 $\pm$ 18 \\
    MCTS & 3511 $\pm$ 56 & 55.6 $\pm$ 1.0 & 992 $\pm$ 8 & 0.04 $\pm$ 0.11 & 498 $\pm$ 18 \\
    Equity-Greedy & 3421 $\pm$ 55 & 49.7 $\pm$ 1.8 & 813 $\pm$ 84 & 0.00 $\pm$ 0.00 & 425 $\pm$ 40 \\
    Equity-MCTS & 3406 $\pm$ 60 & 54.1 $\pm$ 2.1 & 947 $\pm$ 54 & 0.22 $\pm$ 0.39 & 468 $\pm$ 42 \\
    Random & 2068 $\pm$ 289 & 46.4 $\pm$ 3.5 & 590 $\pm$ 128 & 2.95 $\pm$ 0.85 & 264 $\pm$ 70 \\
    Market-Greedy & 1411 $\pm$ 40 & 34.7 $\pm$ 0.0 & 167 $\pm$ 32 & 6.26 $\pm$ 0.10 & 68 $\pm$ 0 \\
    \bottomrule
    \end{tabular*}
    }
    \end{table*}
    
\textbf{Policies.} Every policy is implemented behind a common problem-agnostic interface (valid actions, transition, reward, terminal test, and equity/market scoring hooks) so the \emph{identical} solver code runs on both cases. We evaluate six policies: \emph{Random}; two one-step greedy heuristics: \emph{Equity-Greedy} (target the low-income group's shortfall) and \emph{Market-Greedy} (the naive ``chase the cheap/easy win''); and three planning solvers: receding-horizon \emph{Value Iteration}~\cite{bellman1958dynamic} \rev{with a two-step lookahead}, base \emph{MCTS} (uniform-random rollouts, UCT selection)~\cite{browne2012survey}, and \emph{Equity-MCTS} (MCTS with an equity-biased informed rollout and tree prior). Each policy is evaluated over 30 seeded replications. \rev{The random and greedy rules serve as rule-based benchmarks for these established MDP solvers.} We additionally \emph{cross-deploy} each policy's inductive bias (equity vs.\ market) from one case to the other to study the transferability and \rev{robustness of} the policies.

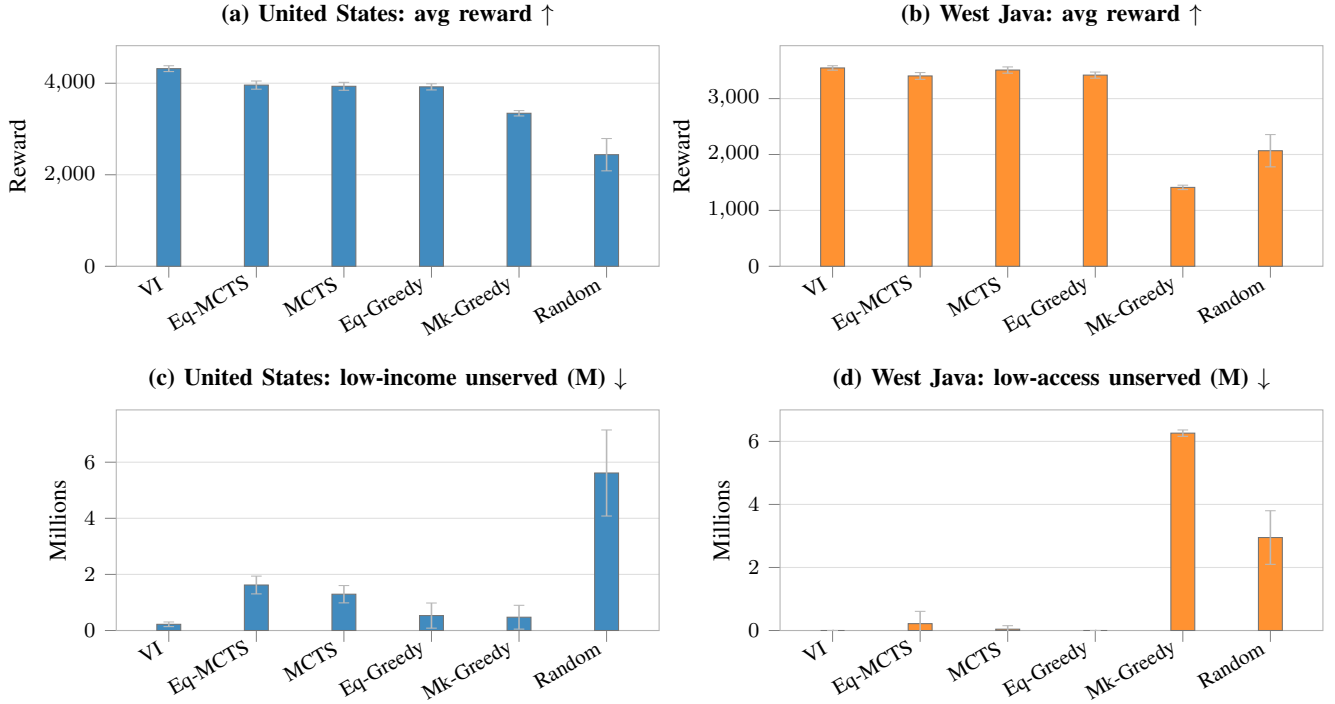
\begin{figure*}[!htbp]
    \centering
\begin{tikzpicture}
\begin{groupplot}[
  group style={group size=2 by 2, horizontal sep=1.6cm, vertical sep=0.9cm, xticklabels at=edge bottom},
  resultbar,
  width=0.5\textwidth, height=3.5cm,
  symbolic x coords={VI,Eq-MCTS,MCTS,Eq-Greedy,Mk-Greedy,Random},
  /pgf/bar width=9pt,
  ymin=0,
]
\nextgroupplot[title={(a) United States: avg reward $\uparrow$}, ylabel={Reward}]
\addplot+[draw=black!55, fill=barUS!85, error bars/.cd, y dir=both, y explicit]
  table[col sep=comma, x=policy, y=us_reward, y error=us_rstd] {figs/data/combined.csv};
\nextgroupplot[title={(b) West Java: avg reward $\uparrow$}, ylabel={Reward}]
\addplot+[draw=black!55, fill=barID!85, error bars/.cd, y dir=both, y explicit]
  table[col sep=comma, x=policy, y=id_reward, y error=id_rstd] {figs/data/combined.csv};
\nextgroupplot[title={(c) United States: low-income unserved (M) $\downarrow$}, ylabel={Millions}]
\addplot+[draw=black!55, fill=barUS!85, error bars/.cd, y dir=both, y explicit]
  table[col sep=comma, x=policy, y=us_unserved, y error=us_ustd] {figs/data/combined.csv};
\nextgroupplot[title={(d) West Java: low-access unserved (M) $\downarrow$}, ylabel={Millions}]
\addplot+[draw=black!55, fill=barID!85, error bars/.cd, y dir=both, y explicit]
  table[col sep=comma, x=policy, y=id_unserved, y error=id_ustd] {figs/data/combined.csv};
\end{groupplot}
\end{tikzpicture}
    \vspace{-2em}
    \caption{Policy comparison: Case~1 U.S. (left, blue) and Case~2 West Java (right, orange). Top: average reward; bottom: priority-group population left underserved (low-income cities in the U.S., low-access regions in West Java). Error bars are 1 std over 30 replications. The market-chasing heuristic (Mk-Greedy) is \rev{sub-optimal} in Case~1 but catastrophic in Case~2.}\label{fig:combined}
    \end{figure*}

    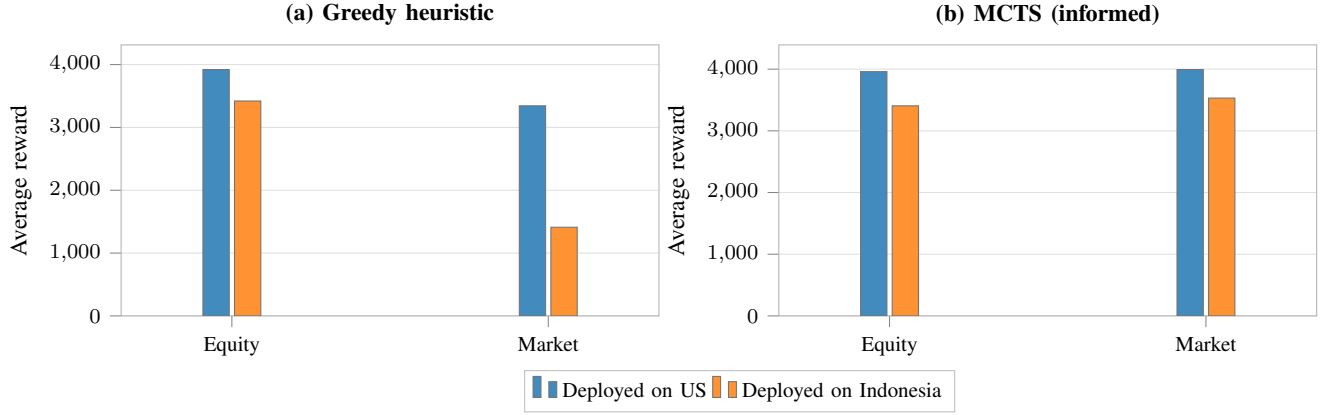
\begin{figure}[!t]
        \centering
\begin{tikzpicture}
\begin{groupplot}[
  group style={group size=2 by 1, horizontal sep=0.3cm, yticklabels at=edge left, ylabels at=edge left},
  groupedbar,
  scale only axis, width=0.36\linewidth, height=2.2cm,
  symbolic x coords={Equity,Market},
  ymin=0, ymax=4500, ylabel={Average reward},
]
\nextgroupplot[title={(a) Greedy heuristic},
  legend style={font=\footnotesize, draw=gray!50, legend columns=2,
                at={(1.04,0)}, yshift=-0.72cm, anchor=north,
                /tikz/every even column/.append style={column sep=6pt}}]
\addplot+[draw=black!55, fill=barUS!85] table[col sep=comma, x=bias, y=us_reward] {figs/data/cross_greedy.csv};
\addplot+[draw=black!55, fill=barID!85] table[col sep=comma, x=bias, y=id_reward] {figs/data/cross_greedy.csv};
\legend{Deployed on US, Deployed on Indonesia}
\nextgroupplot[title={(b) MCTS (informed)}]
\addplot+[draw=black!55, fill=barUS!85] table[col sep=comma, x=bias, y=us_reward] {figs/data/cross_mcts.csv};
\addplot+[draw=black!55, fill=barID!85] table[col sep=comma, x=bias, y=id_reward] {figs/data/cross_mcts.csv};
\end{groupplot}
\end{tikzpicture}
        \caption{Cross-deployment of the two inductive biases (equity vs.\ market) as (a) \rev{a greedy heuristic and (b)} an informed-rollout MCTS planner, each deployed on both cases. The market bias collapses when deployed on Case~2 (a), while lookahead sustains it (b).}\label{fig:cross}
        \end{figure}

\section{Results and Discussion}\label{sec:discussion}

Table~\ref{tab:performance} (Case~1) and Table~\ref{tab:indonesia} (Case~2) give per-policy metrics, \autoref{fig:combined} places the two cases together, and \autoref{fig:cross} shows the cross-deployment transfer. In \emph{both} cases, \textbf{\rev{value iteration} dominates on reward and equity}: in Case~1 it earns the highest reward ($4319\pm64$) at $66\%$ renewable penetration while leaving only $0.22$M low-income residents underserved (a $96\%$ reduction versus Random); in Case~2 it earns the best reward ($3549$), fully closes the access gap, and secures \rev{the second-highest private leverage ($479$, against $498$ for MCTS)}. The other principled solvers \rev{(the two MCTS variants and Equity-Greedy)} cluster just behind in both. The decisive difference is Market-Greedy: the weakest non-random policy in Case~1 ($3344$, with, by coincidence, few underserved because there the largest raw deficits and the priority group align), it \emph{collapses} in Case~2 to the worst reward ($1411$), worst equity ($6.26$M underserved\rev{, which covers} every low-access region), and lowest leverage ($68$), falling below Random. Three findings follow.

\textbf{First, optimization beats heuristics, and equity need not cost efficiency.} In both cases the value-iteration policy attains the best reward \emph{and} the best equity, with the strong planners and the equity heuristic just behind. That the reward-maximizing policies are also the most equitable challenges the assumed equity--efficiency trade-off in infrastructure planning, echoing evidence that well-targeted green-finance and incentive policies expand access without sacrificing productivity~\cite{gao2025green, xie2026government}.

\textbf{Second, the cost of a naive heuristic is economy-dependent} (\autoref{fig:combined}). The reward \emph{ranking} is nearly identical across cases \rev{under the same solver code,} but the penalty for a \rev{poorly designed} heuristic is not. Market-Greedy is only mildly sub-optimal when the planner builds, because there the largest raw deficits and the priority group coincide; it is catastrophic when the planner incentivizes, because ``chase the attractive market'' and ``serve the underserved'' pull apart once deployment runs through private developers who avoid poor and remote regions. A policy's error is thus not a property of the policy alone but of how the economy couples the planner's lever to the equity objective\rev{, which is the central message of the paper and cannot be seen in either case alone}.

\textbf{Third, robustness comes from the objective or from lookahead} (\autoref{fig:cross}). A greedy \emph{equity} bias is safe \rev{in} both cases; a greedy \emph{market} bias collapses in Case~2 ($1411$, $6.26$M underserved), but adding lookahead rescues it (market-biased MCTS: $3531$, $0.03$M) because multi-step planning sees past the myopic target. The dangerous combination is a myopic policy with a mis-targeted bias; sound incentive design should pair targeting rules with at least short-horizon planning.

\rev{\textbf{Sensitivity to the reward weights.} We rescaled each reward weight by $0.5$ and $1.5$, one at a time (six settings in Case~1, ten in Case~2). Value Iteration keeps the highest reward in all sixteen settings, and Market-Greedy stays below Random in every Case~2 setting. Only the middle policies change order, and the Case~2 lead of Value Iteration over MCTS narrows when the access weight is halved.}

\rev{\textbf{Practical implications.} The Case~2 gap is driven by the quota. Market-Greedy exhausts the on-grid quota of the attractive clusters after four incentives and leaves about 83\% of the budget unspent, whereas Value Iteration pairs every incentive with a quota release in the same low-access region. A provincial energy office can use the model to screen a proposed targeting rule before funds are committed.}

\section{Conclusion}\label{sec:conclusion}

We formulated socially-equitable renewable-energy budget allocation as an MDP and evaluated the same policies across a mature economy where the planner builds (eight U.S. cities) and an emerging economy where the planner incentivizes (West Java, Indonesia). Equity-focused value iteration dominates on both reward and equity in both settings, but the cost of a naive market-chasing heuristic is economy-dependent\rev{. It is} mildly sub-optimal under direct construction, yet worse than random under incentive-based deployment, where it abandons the very communities an 
\rev{For emerging economies, the results suggest that incentives should pay for the additional private capital brought to low-access regions and not for installed capacity. In addition, quota should be released together with the incentive in the targeted regions. Finally, off-grid supply should be funded directly where private appetite is weak. Targeting rules should also be paired with short-horizon planning. This study offers new insights for policy makers by analyzing the MDP policies in different economies to help us design a more equitable renewable energy systems for all.}

\bibliographystyle{IEEEtran}
\bibliography{refs}

\end{document}